\documentclass{article}

\usepackage{authblk}

\usepackage{cite}
\usepackage{amsmath}
\usepackage{amsfonts}
\usepackage{algorithmic}
\usepackage{array}
\usepackage{hyperref}
\usepackage{cancel}
\usepackage{needspace}
\usepackage{wrapfig}

\usepackage{fixltx2e}
\usepackage{graphicx}
\usepackage{url}

\newcommand{\ignore}[1]{}

\begin{document}

\title{On the complexity of graphs (networks) by information content,
       and conditional (mutual) information given other graphs}

\author{ Lloyd Allison }

\affil{
  Faculty of Information Technology, \\
  Clayton Campus, Monash University, Australia 3800
}

\date{}

\maketitle

\begin{abstract}
  TR \#2014/277\footnote{HTML version 2014, pdf (LaTeX) version 4 August 2020}.
  This report concerns the \textit{information content} of a graph,
  optionally conditional on one or more background,
  ``common knowledge'' graphs. It describes an algorithm
  to estimate this information content, and includes
  some examples based on chemical compounds.
  \\
\textit{keywords:} Graph, network, complexity, information content,
                   sex drugs, rock and roll (not).
\end{abstract}

\section{Introduction}
\label{sec:Intro}

The information content of a graph, in bits, equals the minimum length of
a message \cite{BW69}, in an efficient code, that sends the graph from
a transmitter to a receiver -- the optimal compressed size of the graph.
If, as is very often the case, the ``true'' statistical model describing
the population of graphs is unknown, information content can be
estimated by using a plausible model.

Some literature on graphs uses `compression' in a different sense,
for example, ``a `compression' $G^*$ of a graph $G$ ...
(a) $G^*$ is a graph with $m*$ edges, where $m* <$ no. of edges in $G$,
(b) it is comp. easy to convert $G$ into $G^*$ and v.v.'' \cite{FM95}~p.1.
(Note, $G^*$ likely has more Vertices than $G$.)
Such $G$ and $G^*$ contain exactly the same information
because it is ``easy to convert'' one into the other and
back again, so this is not compression in the usual
data compression sense.
However, $G^*$ might be easier, in the sense of more convenient,
to compress well, if not optimally, compared to $G$.
And, being sparser, $G^*$ might also appear subjectively simpler,
particularly when drawn in 2D.

The following sections discuss various issues to do with graphs and
their representations, a prototype algorithm, and an
application to the structures of chemical compounds.

\section{Graphs}
\label{sec:Graphs}

A graph, $G=\langle V, E \rangle$, can be directed or undirected.
Each Edge is optionally labelled with a value of type $et$.
Each Vertex is optionally labelled with a value of type $vt$.
If Vertices are unlabelled, set $et=()$, the trivial type, and
if Edges are unlabelled, set $vt=()$.

Depending on the application:
Vertices may or may not be labelled, and
Edges may or may not be labelled.

\begin{table*}[h]
\centering
\begin{tabular}{ |l||l|l| }
\hline
Graph       &  no (v,v)               &  (v,v) allowed \\
\hline \hline
undirected  & $|V| \times (|V|-1)/2$  &  $|V| \times (|V|+1)/2$ \\
  directed  & $|V| \times (|V|-1)$    &  $|V|^2$ \\
\hline
\end{tabular} \\
max. number of edges
\end{table*}

Unless otherwise specified, the following are not allowed here:
\begin{itemize}
\item self-loops,     $\langle v, v\rangle$, and
\item multiple Edges, $\langle u, v\rangle_1$,
                      $\langle u, v\rangle_2$, ...,
    i.e., multi-graphs.
\end{itemize}


\subsection{Vertex orderings}
\label{sec:vertexOrderings}

To encode a graph, one encodes some representation of the graph,
notably the adjacency \textit{matrix} or the
adjacency \textit{lists} representations.
Since every non-redundant representation contains the same information,
that is \textit{the graph}, an optimal encoding of any representation
must give the same
minimum message length (\href{https://www.cantab.net/users/mmlist/MML/}{MML})
as any other.
However one representation might be easier -- more convenient --
to compress than another.

A representation may contain more information than just the graph,
in particular it may imply a certain distinguishable ordering of
the Vertices.
There are $|V|!$ possible orderings of the Vertices although
not all are necessarily distinguishable.
A Vertex ordering (numbering) is arbitrary -- changing it
does not change the graph -- such a representation contains
$log_2(|V|! / |A|)$ bits of information over and
above the graph itself, where $A$ is the automorphism
group of $G$ \cite{Sta05}; the appendix contains examples.
(Note that $|A|=1$ if every Vertex has a unique label.)

An adjacency matrix representation of a graph implies
a particular ordering of the Vertices.
Any distinguishable ordering could be used,
each corresponding to a different adjacency matrix.
Alternatively, some \textit{canonical}
ordering\footnote{  
  A canonical ordering
  $canon(G)$, of the vertices of $G$ can be used
  to determine if two Graphs, $G$ and $H$, are isomorphic --
  $G \sim H$ iff $canon(G)=canon(H)$.
},
and
adjacency matrix, could be used, for example,
the one with its rows in lexicographically non-descending order.
An optimal encoding of it must somehow take advantage
of that property.
Alternatively, to calculate the information content of the graph,
encode any adjacency matrix and subtract $log_2(|V|! / |A|)$ bits.

\subsection{Adjacency matrix}
\label{sec:adjacency}

An adjacency matrix of a graph can be trivially encoded using $1$~bit
per possible Edge.
This is only optimal if half of the possible Edges exist, and
the existence of an Edge is independent of other Edges, and
of labels if present.

When the Edge density is not known apriori,
the \textit{very least} that should be done is to model the
entries of the adjacency matrix by a binomial distribution.
This is optimal if the existence of an Edge is
independent of other Edges.
The hypothesis has one parameter, the expected Edge density,
which is the same in all parts of the graph.
This does not mean that all Vertices have the same degree.
Note, subtly different is the hypothesis that
the Edge density is homogeneous
(and we do not care what the density is)
which can be realised by using the \textit{adaptive} coding of
binomial data and which saves a fraction of a bit in total.

For a \textit{directed} graph, if it is believed that the degrees
of Vertices may differ and are independent, and the
existence of Edges are independent, an approriate model
is to use one binomial distribution per row (or column)
of the adjacency matrix.
The model has one parameter per Vertex,
being the out- (or in-) Edge density of the Vertex.
(Many simple variations are possible, for example,
a model could have two density parameters, for
``low'' and ``high'' degree vertices, plus a
binomial distribution over into
which of these ``classes'' Vertices fall.)

The situation is more complicated for an \textit{undirected} graph where
an adjacency matrix is necessarily symmetric (equivalently upper
or lower triangular).
The existence of an undirected Edge $(u,v)$ depends on
the neighbourliness of both $u$ and $v$.
These cannot be independent:
consider ``$u$ is adjacant to all, and $v$ is adjacent to none.''
It is tempting to average the contributions of $u$ and $v$.

\subsection{Special classes of graph}
\label{special}

Methods have been devised to efficiently encode some
special classes of graph:
\begin{itemize}
\item rooted strict binary tree, about $1$-bit per
  Vertex\footnote{  
    A (rooted) strict binary tree, where every node either has
    two subtrees (is a Fork), or has zero subtrees (is a Leaf),
    can be encoded in 1-bit per vertex (node):
    Perform a prefix traversal of the tree, output `F' for a Fork,
    `L' for a Leaf. The end can be detected when \#L=\#F+1.
    (This preserves the left-to-right ordering of subtrees;
    savings are possible if this is unnecessary.)
  },
\item rooted general tree, about $2$ bits per
  Edge\footnote{  
    A (rooted) general tree can be encoded in about 2-bits per edge:
    Perform a prefix traversal of the tree, output `d', for down,
    when descending an edge, and output `u', for up,
    when returning up an edge.
    The end can be encoded as an extra `u', say.
    (This preserves the left-to-right ordering of subtrees;
    savings are possible if this is unnecessary.)
  },
\item plane graph, about $4$ bits per Edge \cite{Tur84},
    later $log_2(14)$ bits per Edge \cite{Via08}.
\end{itemize}

The above examples rely on global properties of a graph and
do not attempt to take advantage of any kind of repeated pattern.

More interesting are attempts to base graph compression
on recognising patterns (motifs), that are repeated,
possibly approximately, that is with variations.
This is particularly so where a graph has \textit{semantics} --
where its structure carries meaning.

\section{A prototype algorithm}
\label{sec:prototype}

The idea is to extend the Lempel-Ziv style of
sequence compression to graphs:
traverse the graph in some arbitrary order.
At each step in the traversal there is the next element,
the next \textit{unknown} -- a Vertex or an Edge.
Seek scored matches to the recently traversed, and
hence known, \textit{context} in the common knowledge,
on the basis that what happened ``then'' might happen ``now.''
Form a probability distribution over the possible values
for the next element (Vertex (label, arity), Edge (label, looping))
using the scored (weighted) matches.
The common knowledge consists of zero or more given graphs.
(For simplicity, the prototype does not search for matches
to the context within the traversed part of the graph
being compressed, but this is clearly possible in principle.)

The method described below calculates the information
content of an undirected graph.
It is clear that, in principle, a ``similar method''
can be devised for directed graphs.
(The probability distribution calculated at each
step could be used, with an arithmetic compressor, say,
to perform actual data compression.)

The prototype does not attempt to ``recover'' the extra
information from the arbitrary Vertex ordering
(see automorphism group earlier).

It is sufficient to consider a single connected component;
a disconnected graph can be treated one component at a time,
including earlier components in the common knowledge, if desired.

\begin{verbatim}
data Vertex vt et = Vertex vt Int [Edge vt et]
data Edge vt et = Edge et (Vertex vt et) (Vertex vt et)
  -- a Vertex contains an arbitrary but unique Int,
  -- only to identify it; this is not a "label".

-- e.g., ...
data Entity     = Utility | House
                  deriving (Eq, Enum, Bounded, Show, Read)
data Connection = Elec | Gas | H2O
                  deriving (Eq, Enum, Bounded, Show, Read)

k33 = v00 where                  -- complete, bipartite, labelled
  v00 = Vertex Utility 0
        [Edge Elec v00 v10, Edge Elec v00 v11, Edge Elec v00 v12]
  v01 = Vertex Utility 1
        [Edge Gas v01 v10, Edge Gas v01 v11, Edge Gas v01 v12]
  v02 = Vertex Utility 2
        [Edge H2O v02 v10, Edge H2O v02 v11, Edge H2O v02 v12]

  v10 = Vertex House 3
        [Edge Elec v10 v00, Edge Gas v10 v01, Edge H2O v10 v02]
  v11 = Vertex House 4
        [Edge Elec v11 v00, Edge Gas v11 v01, Edge H2O v11 v02]
  v12 = Vertex House 5
        [Edge Elec v12 v00, Edge Gas v12 v01, Edge H2O v12 v02]
\end{verbatim}


\subsection{Traversal of an undirected graph}
\label{undirected}

An \textit{undirected graph}
is represented by an adjacency-lists data structure;
if Vertices u and v are adjacent, the data structure links
$u$ to $v$ and also links $v$ to $u$.

The information-content calculations rely on a graph traversal.
The traversal arrives at a Vertex via an incoming Edge and
follows an Edge out from a Vertex, so the
\textit{immediate} context of an Edge is a Vertex, and
the immediate context of a Vertex is an Edge
(except for the initial Vertex which has the empty context).

\begin{verbatim}
traverse_undirected fv fe v =
  ...
\end{verbatim}

\begin{wrapfigure}{R}{.52\textwidth}
\centering
\includegraphics[width=0.5\textwidth]{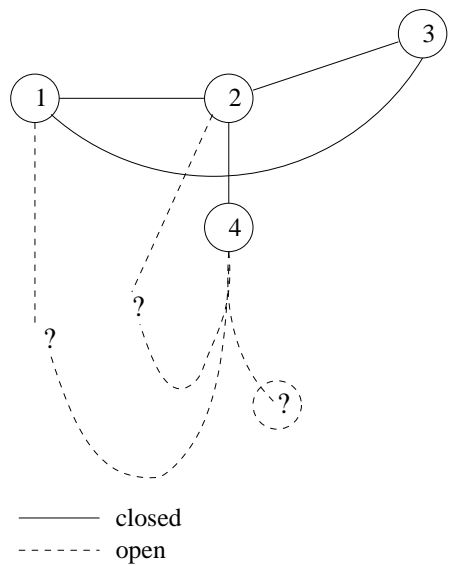}
\caption{Vertex traversal order: 1, 2, 3, 4.
         Adjacency list order ``clockwise''.
         Vertices 1, 2, 4 VISITING; Vertex 3 VISITED.}
\label{fig:trav}
\end{wrapfigure}

Traverse a graph accessible from Vertex $v$,
apply $fv$ to each Vertex and $fe$ to each Edge, and
return a list of type \texttt{ [Either fv\_result fe\_result] }.
The output contains one element per Vertex and one element per Edge.

During traversal, a Vertex has a `status',
one of \texttt{UNVISITED | VISITING | VISITED}
(see figure \ref{fig:trav}).
Initially every Vertex is \texttt{UNVISITED}.
A Vertex becomes \texttt{VISITING} when the traversal
first accesses it, and remains so while any of
its Edges are yet to be traversed.
A Vertex becomes \texttt{VISITED} when all its Edges have been traversed.

An Edge involving a \texttt{VISITING} Vertex is
either `open' (untraversed) or `closed' (traversed).
Note, an Edge is necessarily closed if both of
its Vertices are \texttt{VISITED}.
(Nothing at all is known of an Edge both of
whose Vertices are \texttt{UNVISITED}.)

The traversal maintains a list of \texttt{VISITING} Vertices
and a list of \texttt{VISITED} Vertices, which are
passed to $fv$ and $fe$; $fv$ is also given the
optional (reversed) incoming edge to the Vertex.
The \texttt{VISITING} list acts as a stack during traversal and
when a Vertex is popped it becomes \texttt{VISITED}.

\subsection{Information on Vertices and Edges}

\begin{verbatim}
predict  vertexLabel2maxDegree v1 v2s =
  let
    unwrap (Left(Vertex vLabel _ ves, scored)) =
      ...
    unwrap (Right(Edge eLabel _ v2, loop_candidates, scored)) =
      ...
  in map unwrap (traverse_and_match v1 v2s)
\end{verbatim}

For the graph accessible from Vertex $v1$,
calculate the negative log probability of
\begin{enumerate}
  \item the label and degree of each Vertex, and
  \item the label and possible loop-completion of each Edge
\end{enumerate}
given graphs accessible from Vertices $v2s$.
To do this, traverse the graph accessible from $v1$, using $v2s$ as
background knowledge.
The traversal returns a list of scored context matches for
each Vertex, and similar for each Edge.
A list of scored matches is converted into
a Model (the prediction), which gives the
negative log probability ($nlPr$) of the Vertex (or Edge) information.

\begin{verbatim}
traverse_and_match  v1 v2s =
  let given_Vs = concat(map vertex_list v2s)
  in traverse_undirected (v_matches given_Vs) (e_matches given_Vs) v1
\end{verbatim}

\noindent
\texttt{traverse\_undirected} (see earlier) is a
general purpose traversal function, taking a function
to apply to each Vertex, a function to apply to
each Edge, and the ``root'' of a graph.

\subsection{A Vertex has an Edge for immediate context}

\begin{verbatim}
v_matches  v2s visiting visited back_e v1
  -- Return scored matches to (v1's) context from amongst
  -- the v2s.  v1 is as yet unknown during a traversal;
  -- we are trying to predict it.   back_e is our "context"
  -- [e1], [v1--e1-->prev_v],  unless it is [].
    ...
    v20 @ (Vertex _ _ v20es) <- v2s,
    e2 <- v20es,
    let [e1] = back_e
        (_, score) = match_Edge ... ... ... e1 e2,
    ...
\end{verbatim}

During traversal one at arrives at a Vertex by an Edge,
\texttt{back\_e} (reversed), the most recent part of the `context'.
(The initial Vertex is an exception, having no context, \texttt{[]}.)
Any Vertex, $v2$, in the given $v2s$ is a potential prediction
for $v1$, with a score depending on how much graph around $v2$
matches the context.

\subsection{And an Edge has a Vertex for immediate context}

\begin{verbatim}
e_matches  v2s visiting visited (e1 @ (Edge e1l v10 _ )) =
  -- Return scored matches to (e1 and its) context from amongst
  -- the v2s' Edges. e1 (v10--e1-->v11) and v11 are as yet unknown
  -- during the traversal, except that v10 is known "context".
  -- We are trying to predict e1 including whether it closes a
  -- or not; we cannot and do not use actual knowledge of v11.
    ...
    v20 @ (Vertex _ _ v20es) <- v2s,
    e2 @ (Edge _ _ v21) <- v20es,
    match_Vertex ... ... ... v10 [e1] v20 [e2]
\end{verbatim}

Similarly, during traversal one arrives at an Edge, $e1$,
from a Vertex, $v10$, the most recent part of the Edge's ``context''.
Any Edge $e2$, in (the graphs accessible from) $v2s$ is a
potential prediction for $e1$, with a score depending on
how much graph around $e2$ matches the context.

Note that \texttt{UNVISITED} Vertices and open Edges of
the graph being traversed are as yet unknown and cannot
be used in matching and scoring, although the existence
of an open Edge (but nothing more) is known.

\subsection{Matches and Scores}

\begin{verbatim}
match_Vertex pv pe corr (v1 @ (Vertex _ _ v1e)) back_e1
                        (v2 @ (Vertex _ _ v2e)) back_e2 =
  ...
\end{verbatim}

Calculate how much of the graph rooted at $v1$ matches
the graph rooted at $v2$.
This could be done in a great many ways depending on
the application.
At the simplest, $v1$ and $v2$ must have the same label,
as a start.
Beyond that, $v1$'s Edges (and beyond) and \textit{some permutation}
of $v2$'s Edges (and beyond) may match.

\begin{verbatim}
match_Edge pv pe corr (e1 @ (Edge _ _ v11)) (e2 @ (Edge _ _ v21)) =
  ...
\end{verbatim}

At the simplest, Edges $e1$ and $e2$ must have the same label,
 as a start.
Beyond that, their ``to'' Vertices (and beyond) may match.

Note that an open Edge can match any Edge in the
background graphs, but it contributes zero to the score,
and matching does not continue beyond it.

\subsection{Example}

\begin{verbatim}
I(k33|[]) = 47.0
I(k33|[k33]) = 23.4
I(k33|[near_k33]) = 37.3
where k33 is as above and

near_k33 = v00 where
  v00 = Vertex Utility 0
        [Edge Elec v00 v10, Edge Elec v00 v11, Edge Elec v00 v12]
  v01 = Vertex Utility 1
        [Edge Gas  v01 v10, Edge Gas  v01 v12, Edge Gas  v01 v13]
  v02 = Vertex Utility 2
        [Edge H2O  v02 v10, Edge H2O  v02 v11, Edge H2O  v02 v12,
        Edge H2O  v02 v13]
  v10 = Vertex House 3
        [Edge Elec v10 v00, Edge Gas v10 v01, Edge H2O v10 v02]
  v11 = Vertex House 4
        [Edge Elec v11 v00, Edge H2O v11 v02]
  v12 = Vertex House 5
        [Edge Elec v12 v00, Edge Gas v12 v01, Edge H2O v12 v02]
  v13 = Vertex House 6 [Edge Gas v13 v01, Edge H2O v13 v02]
\end{verbatim}

\begin{figure}
\centering
\small{Carbon labels are generally omitted in renderings.}
\begin{tabular}{ ll }
\includegraphics[width=0.9\textwidth]{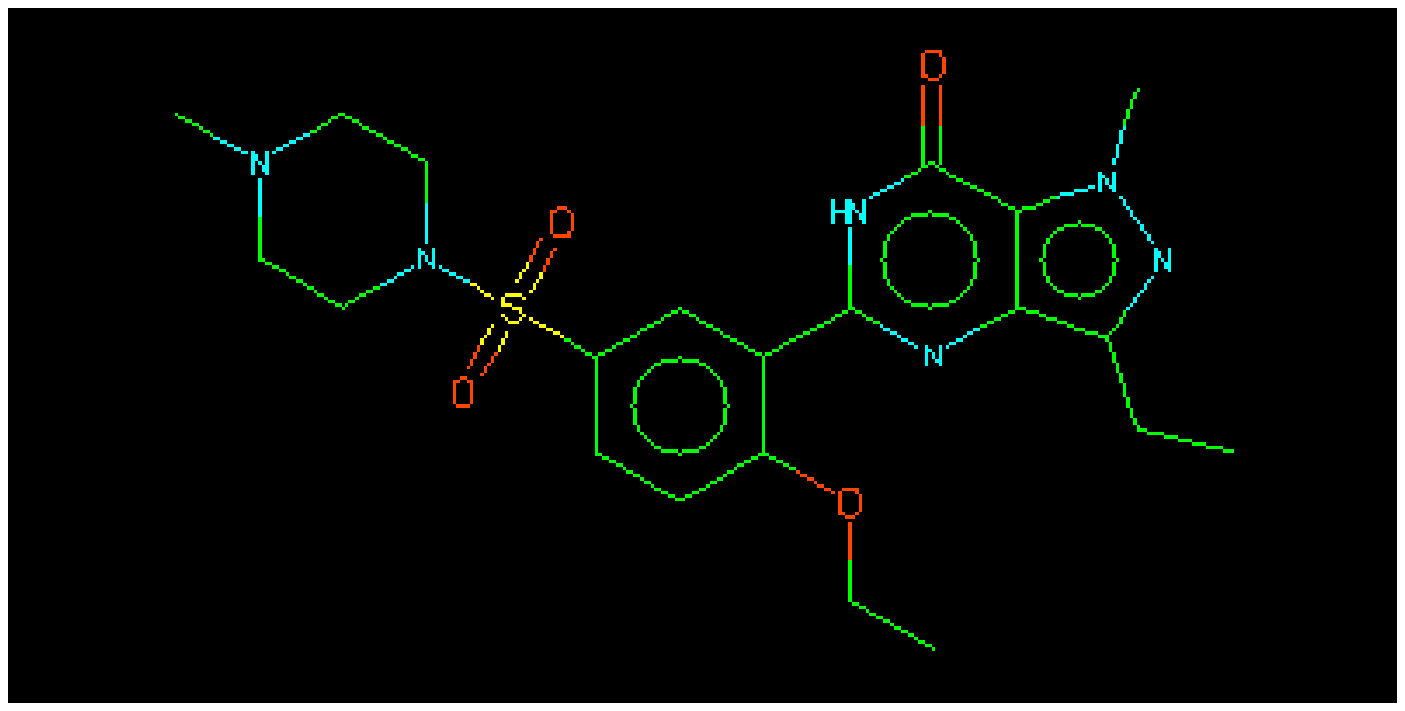}  & viagra  \\
\includegraphics[width=0.9\textwidth]{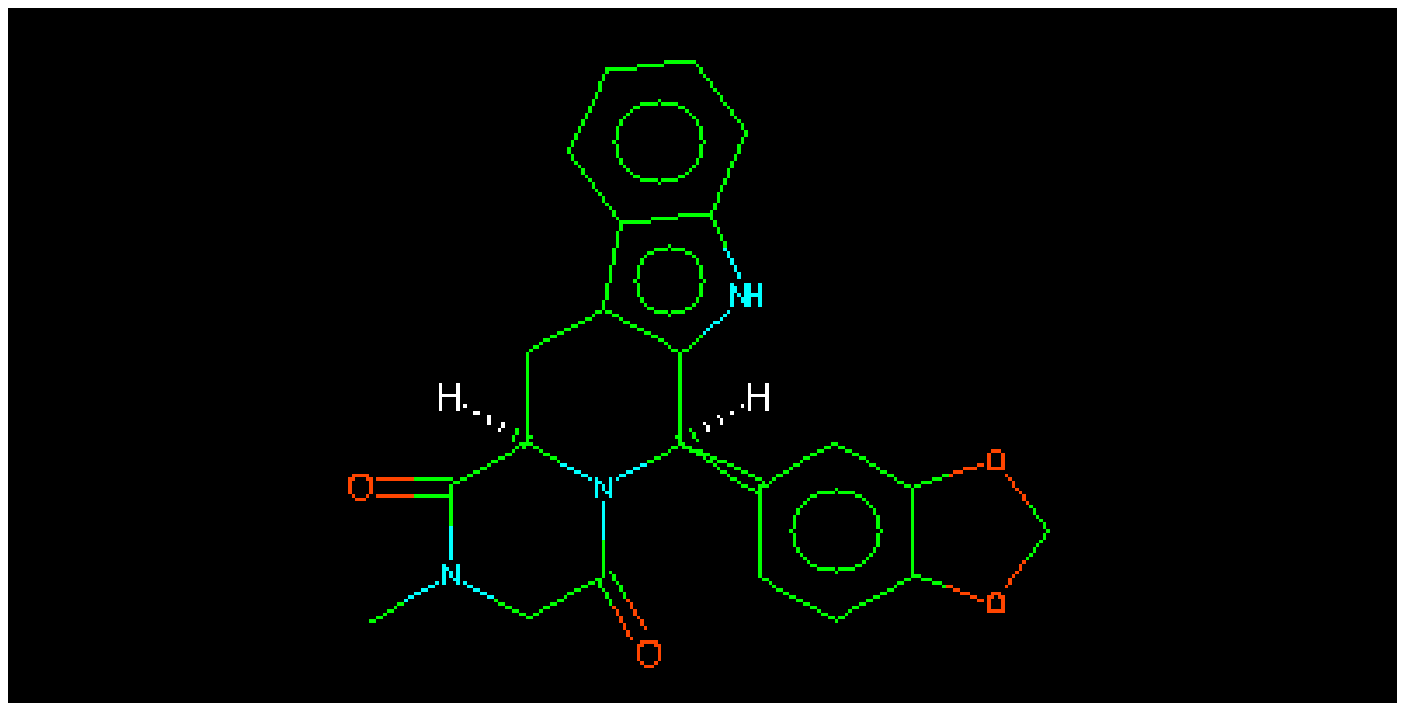}  & cialis  \\
\includegraphics[width=0.9\textwidth]{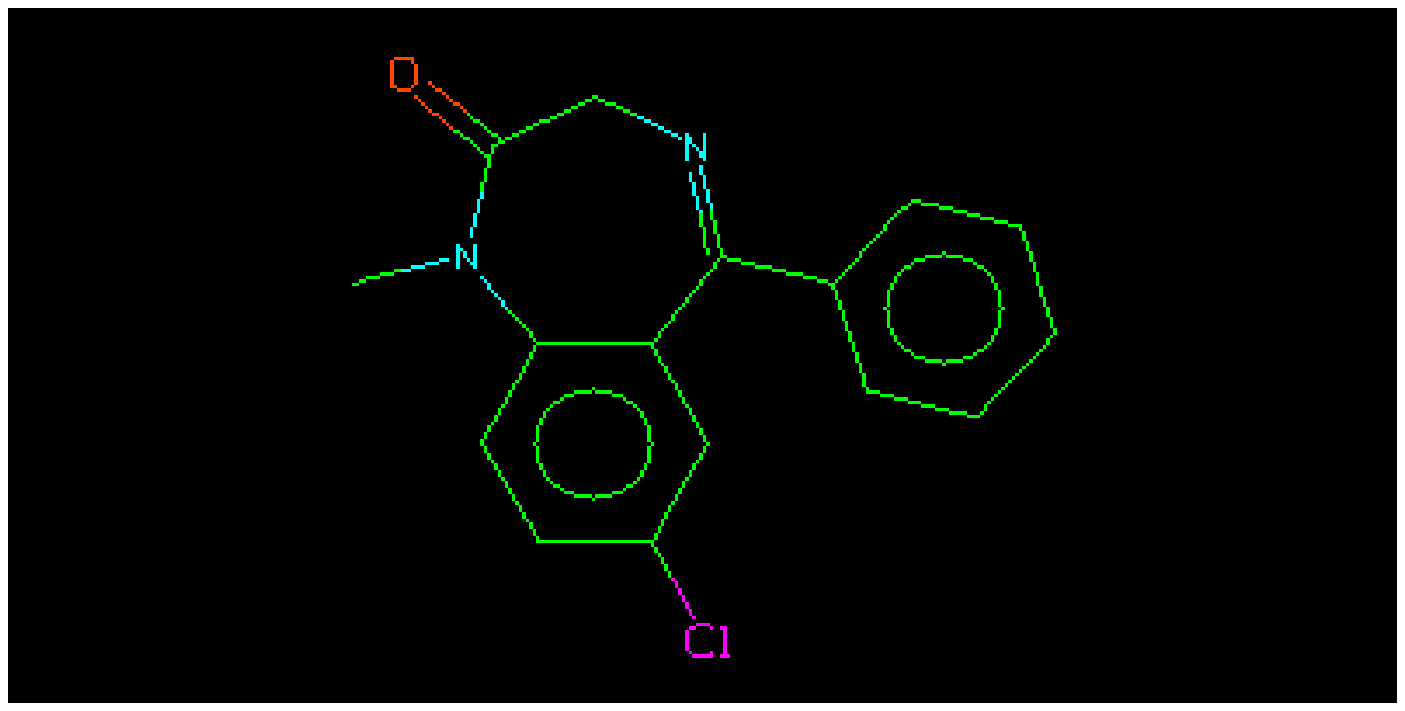}  & valium  \\
\includegraphics[width=0.9\textwidth]{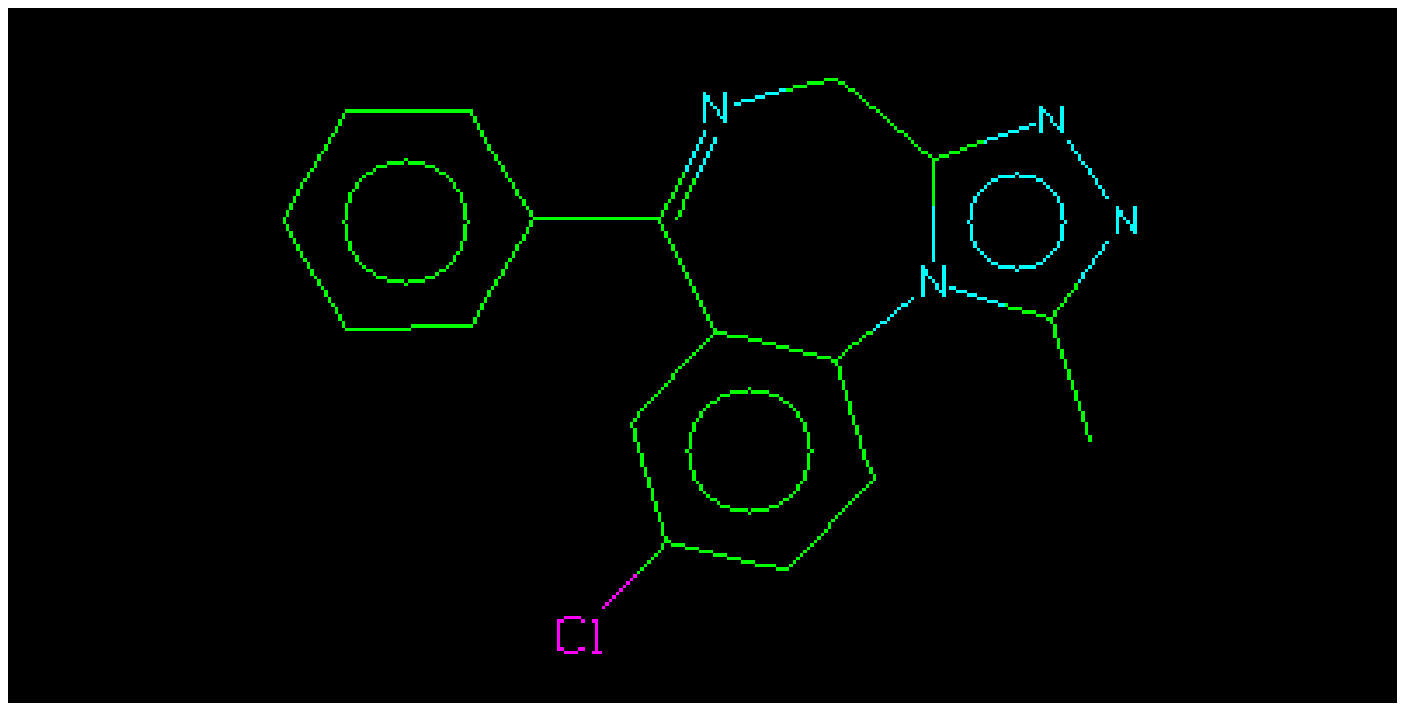}   & xanax   \\
\end{tabular}
\caption{ (Examples at daylight.com) }
\label{fig:f0}
\end{figure}

\section{Application to chemical compounds}

For present purposes, a chemical compound is a graph where
each Vertex is labelled with a chemical element, and
each Edge is labelled with a bond.
SMILES \cite{Wei88} is a language for defining chemical compounds.
Some functions were written to manipulate SMILES:

\begin{verbatim}
data Atom = ... Element ...
data Exp  = ... -- type of SMILES parse trees
instance Read Exp where ... -- i.e., parser for SMILES expressions
\end{verbatim}
the following are sufficient for many biochemical molecules...
\begin{verbatim}
data Element = Hydrogen | Carbon | Nitrogen | Oxygen
             | Phosphorous | Sulphur | Bromine | Chlorine
             | Iodine
data Bond = No_bond | Sng_bond | Dbl_bond | Trp_bond | Arm_bond
 . . .
smiles2Graph: return the graph corr. to a given SMILES parse tree,
infer_implicit_bonds: "obvious" bonds can be omitted from
                      SMILES expressions,
atoms2Elements: extract the element in each atom (SMILES allows
                the specification of other information such as
                charge and chirality, not used here).
 . . .
viagra = "CCc1nn(C)c2c(=O)[nH]c(nc12)c3cc(ccc3OCC)S(=O)(=O)N4CCN(C)CC4"
cialis = "CN1CC(=O)N2[C@@H](c3[nH]c4ccccc4c3C[C@@H]2C1=O)c5ccc6OCOc6c5"
valium = "CN1C(=O)CN=C(c2ccccc2)c3cc(Cl)ccc13"
xanax  = "Cc1nnc2CN=C(c3ccccc3)c4cc(Cl)ccc4-n12"
\end{verbatim}
For examples see figure~\ref{fig:f0}.

Chemical compounds are of bounded degree.
Most small compounds are planar but exceptions are conceivable.
Hydrogen atoms can be specified in SMILES but they are
usually omitted and, if so, they, and the bonds to them,
are left out of the resulting graph.

The information-content routines were ``told'' that the
graphs were of bounded degree, and were given the maximum degree
of each element.
Apart from that, they were not given any hard-wired chemical information.
It is of course possible to include example graphs as background knowledge.

The following table shows the results of experiments on viagra,
cialis, valium and xanax [daylight.com],
singly and in various combinations.

\begin{table*}[h]
\centering
\begin{tabular}{ |l|l|l|l|l| }
\hline
viagra & cialis & valium & xanax & total \\
\hline
267    & 262    & 169    & 194   & 892  \\
\hline
\end{tabular} \\
information(.) in bits
\end{table*}

\begin{table*}[h]
\centering
\begin{tabular}{ |l|l|l|l|l| }
\hline
       & viagra & cialis & valium & xanax  \\
\hline
viagra & (75)   & 180    & 217    & 211    \\
cialis & 168    & (55)   & 160    & 172    \\
valium & 119    & 108    & (48)   & 86     \\
xanax  & 128    & 133    & 115    & (59)   \\
\hline
\end{tabular} \\
information(left$|$top) in bits\\
\small{(The diagonal elements are small, but not tiny,
       because there is no great prior expectation of X$|$X.)}
\end{table*}

\begin{table*}[h]
\centering
\begin{tabular}{ |l|l|l|l|l| }
\hline
viagra & cialis$|$viagra & valium$|$viagra, & xanax$|$valium, & total   \\
       &                 & ~~cialis         & ~~viagra,cialis & \\
\hline
267 & 168 & 111 & 88 & 634 \\
\hline
\end{tabular} \\
information(.$|$.) in bits \\
  \small{(Note that viagra \textit{and} cialis are less
  helpful to valium than cialis alone.)}
\end{table*}


\section{Summary}

The problem of calculating a graph's information content is so
general as to be almost insufficiently specified when one
considers factors such as but not limited to:\\
\begin{center}
\begin{tabular}{ |l|l| }
\hline
degree:             &  bounded$|$unbounded, sparse$|$dense  \\
\hline
(in?)dependence of: & vertex label, Vertex degree,  \\
                    & Edge label, cycle length \\
\hline
label matching:
   & equal or ``close''  (perhaps  \\
   & squared difference for continuous?) \\
\hline
subgraph matching:
   & equal or approximate (even varieties of \\
   & approximate tree matching are hard) \\
\hline
\end{tabular}
\end{center}

A prototype algorithm exists for some of the above options ``pinned down.''
It has been used to calculate the information content of
graphs derived from small chemical compounds.
It takes a few seconds to process compounds containing
dozens of elements.
Its worst-case time complexity is exponential,
but certain speed-up techniques are conceivable, such as,
\begin{itemize}
    \item hash table of small sub-graphs to find good
          candidate matches quickly, and
    \item halting context matching at a distance threshold.
\end{itemize}



\bibliographystyle{plainurl}

\bibliography{paper}

\noindent{
(Added later: Also see
L. Allison,
`\textit{Coding Ockham's Razor}',
Springer, 2018, isbn13:~978{-}3319764320,
\href{https://doi.org/10.1007/978-3-319-76433-7}
                 {doi:10.1007/978-3-319-76433-7},
and\newline
\href{https://www.cantab.net/users/mmlist/MML/}
             {www.cantab.net/users/mmlist/MML/}}.)


\raggedbottom
\needspace{9\baselineskip}
\section{Appendix}
\label{appendix}

Examples of graphs, automorphism groups and distinguishable Vertex orderings.

\begin{verbatim}
K4  *----*   |A|=4!
    |\  /|
    | \/ |   all 4! Vertex orderings are indistinguishable
    | /\ |
    |/  \|
    *----*
\end{verbatim}
ditto the completely disconnected graph.
\needspace{4\baselineskip}
\begin{verbatim}
C4  *--*     |A|=8  (4 rotations x reflection)
    |  |
    |  |
    *--*
\end{verbatim}

\needspace{5\baselineskip}
\begin{verbatim}
  4!/8 = 3 distinguishable Vertex orderings
    1--2    1--2    1--3
    |  |    |  |    |  |
    |  |    |  |    |  |
    4--3    3--4    4--2
\end{verbatim}
and their implied adjacency matrices
\needspace{4\baselineskip}
\begin{verbatim}
    _101    _110    _011
    1_10    1_01    0_11
    01_1    10_1    11_0
    101_    011_    110_
\end{verbatim}

\needspace{4\baselineskip}
\begin{verbatim}
G = *--*    |A|=4   (reflection x reflection)
    |\ |
    | \|
    *--*
\end{verbatim}

\needspace{5\baselineskip}
\begin{verbatim}
  4!/4 = 6 distinguishable Vertex orderings
    1--2    1--2    1--3    2--1    2--1    3--1
    |\ |    |\ |    |\ |    |\ |    |\ |    |\ |
    | \|    | \|    | \|    | \|    | \|    | \|
    4--3    3--4    4--2    4--3    3--4    2--4
\end{verbatim}
Note that the following Vertex-labelled graph
\needspace{5\baselineskip}
\begin{verbatim}
G = W--B
    |  |
    |  |
    B--W
\end{verbatim}
also has the same $|A|=4$.

\end{document}